\documentclass[journal,twoside,web]{ieeecolor}
\usepackage{generic}
\usepackage{amsmath,amssymb,amsfonts}
\usepackage{algorithmic}
\usepackage{graphicx}
    \graphicspath{ {./images/} }
\usepackage{textcomp}
\def\BibTeX{{\rm B\kern-.05em{\sc i\kern-.025em b}\kern-.08em
    T\kern-.1667em\lower.7ex\hbox{E}\kern-.125emX}}

\usepackage{multicol}
\usepackage{multirow}
\usepackage{tabularx}              
\usepackage{booktabs}
\usepackage{hyperref}                
  \hypersetup{hidelinks,colorlinks=false}  
\usepackage{placeins}

\newcommand{\sus}[1]{$^{\mbox{\scriptsize #1}}$}

\newcommand{\fig}[1]{Fig.~\ref{#1}}
\newcommand{\tab}[1]{Table~\ref{#1}}

\newcommand{\sect}[1]{Section~\ref{#1}}

\newcolumntype{Z}{>{\centering\arraybackslash}X}  

\usepackage[backend=biber,style=ieee,maxbibnames=3,bibencoding=utf8]{biblatex}   
\begin{document}

\title{Design and Validation of a Portable EEG–tES Platform Supporting High-Rate EEG Recording and Temporal Interference Stimulation}


\author{
Le~Xing\sus{1,2,3,*},
Maoxing Liang\sus{3},
Disi~A\sus{3,*},
Yifei Jia\sus{3},
Gaoxiang Xie\sus{3},
Linfeng Xu\sus{3},
Xiang'ao Chen\sus{3},
Jiebin Shi\sus{3},
Gang~Pan\sus{1,2},
and~Bicheng~Han\sus{3}
\thanks{\sus{1}College of Computer Science and Technology, Zhejiang University, Hangzhou, China.}
\thanks{\sus{2}State Key Laboratory of Brain-machine Intelligence, Zhejiang University, Hangzhou, China.}
\thanks{\sus{3}Zhejiang Qiangnao Technology Co., Ltd.}
\thanks{\sus{*}Corresponding author. Emails: le.xing@brainco.cn; adis@brainco.tech}
}

\maketitle

\begin{abstract}
    \textbf{Background:} Closed-loop neuromodulation integrating electroencephalography (EEG) and transcranial electrical stimulation (tES) has strong potential for neuroscience research and clinical applications. However, existing platforms often rely on benchtop instrumentation or FPGA-centered architectures, limiting portability and increasing system complexity. \textbf{Methods:} We developed a compact wearable bidirectional platform integrating 8-channel EEG acquisition and 2-channel tES within a single microcontroller unit (MCU). An ADS1299-based front-end supports up to 8 kHz sampling per channel for simultaneous neural and stimulation signal recording. The stimulation module uses direct digital synthesis (DDS) technique to generate programmable tDCS, tACS, and tTIS waveforms, with optimized firmware design enabling real-time operation on the microcontroller. \textbf{Results:} Experimental evaluation demonstrated high signal fidelity, with EEG correlation coefficients reaching $99\%$ under controlled conditions and $93.5\%$ on a gelatine head phantom. Stimulation performance showed current errors below $1\%$ across tDCS, tACS, and tTIS modes. The platform also reliably recorded concurrent EEG--tES signals without saturation during high-amplitude and high-frequency stimulation. \textbf{Conclusion:} These findings demonstrate that an MCU-centered architecture can effectively support simultaneous EEG sensing and multimodal tES delivery in a compact wearable form factor, while maintaining accurate stimulation and robust signal acquisition. \textbf{Significance:} This work provides a practical foundation for portable closed-loop neuromodulation systems with integrated host software, expanding access to personalized stimulation paradigms, and future point-of-care neurotechnology applications.
\end{abstract}


\section{Introduction} \label{sec:intro}

Noninvasive neuromodulation has become an important tool in both fundamental neuroscience and emerging clinical therapy \cite{rektorova2025non,to2018changing}. Among the available approaches, the combination of wearable electroencephalography (EEG) and transcranial electrical stimulation (tES) has attracted increasing interest because it enables bidirectional interaction with the brain: EEG provides millisecond-scale monitoring of neural dynamics, while tES modulates cortical excitability through externally applied electric fields \cite{miniussi2012combining,arpaia2025combined}. Integrating these two modalities within a single platform creates the foundation for closed-loop neuromodulation systems, in which stimulation parameters can be adjusted in real time according to ongoing brain activity. Such systems have the potential to improve therapeutic precision, enhance mechanistic understanding of brain--stimulation interactions, and support personalized interventions \cite{soleimani2023closing,arpaia2025combined}.

Different tES waveforms provide complementary neuromodulatory mechanisms. Transcranial Direct Current Stimulation (tDCS) modulates cortical excitability through subthreshold shifts in membrane polarization and has been widely investigated for cognitive enhancement and treatment of neuropsychiatric disorders \cite{narmashiri2025effects,zheng2024evaluating}. Transcranial alternating current stimulation (tACS) delivers oscillatory currents that can entrain endogenous neural rhythms, making it valuable for studying functional connectivity and rhythm-specific modulation \cite{elyamany2021transcranial}. Transcranial random noise stimulation (tRNS) applies broadband stochastic currents and has been associated with increased cortical responsiveness \cite{rektorova2025non}. More recently, transcranial temporal interference stimulation (tTIS), first introduced by Grossman \textit{et al.} \cite{grossman2017noninvasive}, has attracted substantial attention because it uses two kilohertz-range carrier signals to generate a low-frequency interference envelope in deeper brain regions. Compared with conventional tDCS and tACS, which primarily affect superficial cortical areas, tTIS offers the possibility of more focal and deeper neuromodulation while reducing direct stimulation of overlying tissue \cite{mirzakhalili2020biophysics,neudorfer2021kilohertz,wang2025temporal}. Owing to these advantages, tTIS has been actively explored for applications including epilepsy, motor control, and memory enhancement \cite{violante2023non,acerbo2022focal,esmaeilpour2021temporal,xu2025precision,vassiliadis2024safety}.

Despite rapid progress in stimulation paradigms, hardware platforms for closed-loop EEG--tES research remain limited. Many current systems still rely on separate laboratory-grade stimulators and EEG amplifiers, increasing system complexity and hindering real-time synchronization, portability, and deployment outside laboratory environments. For example, the neuroConn LOOP-IT platform (neurocare Ltd., Germany) provides precise closed-loop stimulation with broad modality support, including tTIS, but relies on host-computer-assisted instrumentation with a relatively large benchtop footprint \cite{neuroConn}. In contrast, the Neuroelectrics Starstim platform (Neuroelectrics Ltd., Spain) integrates EEG and stimulation into a wearable form factor, but supports only tDCS, low-frequency tACS, and tRNS, and therefore cannot generate the kilohertz carrier frequencies required for tTIS \cite{starstim, ruffini2015application}. In addition, according to publicly available specifications, mainstream integrated EEG--tES devices typically provide EEG sampling rates of no more than 1~kHz, which is insufficient for directly recording high-frequency stimulation waveforms or temporal interference carrier signals during concurrent operation.

Beyond commercial devices, many research prototypes also face practical tradeoffs. Several systems rely on FPGA- or ASIC-based architectures to achieve high throughput and artifact suppression. For example, the WAND platform demonstrated advanced closed-loop neuromodulation performance, but its architecture remains relatively power-intensive and less suitable for compact wearable use \cite{zhou2017wand}. Other studies have demonstrated MCU-based stimulators for tTIS generation \cite{zhang2022designing}, yet these systems were not fully integrated with multichannel EEG acquisition and remained relatively bulky prototypes. Consequently, a key unmet need remains for a compact, low-cost, and fully integrated platform that simultaneously provides: 1) high-rate multichannel EEG acquisition, 2) precise and programmable multi-modal stimulation, 3) support for kilohertz-range tTIS generation and recording, and 4) portable operation suitable for future wearable closed-loop applications. In parallel with advances in neuromodulation, wearable EEG systems have highlighted the value of portable neural monitoring beyond laboratory settings, including home-based, ambulatory, and long-duration applications. However, comparable progress in integrated EEG--tES hardware remains limited, particularly for systems requiring high-rate sensing and flexible stimulation. Extending wearable design principles to bidirectional EEG--tES platforms could broaden access to personalized real-world closed-loop neuromodulation \cite{casson2019wearable,niso2023wireless,soleimani2023closing}.

Developing such a system is technically challenging. High-speed EEG acquisition must preserve microvolt-level neural activity while tolerating volt-level stimulation artifacts, particularly during concurrent tTIS where carrier signals occur in the kilohertz range. At the same time, the stimulator must deliver accurate current-controlled outputs with low distortion, flexible waveform programmability, and minimal interference to the sensing module. Achieving these functions within stringent constraints of power consumption, computational resources, size, and cost is a major engineering barrier.

In this work, we present a portable bidirectional EEG--tES platform that addresses these challenges using a microcontroller-centered architecture. The proposed system integrates an STM32 MCU, an 8-channel EEG front-end, and a custom high-fidelity stimulation module based on DAC and direct digital synthesis (DDS) techniques. The device supports multiple stimulation modalities including tDCS, tACS, tTIS, and programmable ramping profiles, while enabling simultaneous EEG recording at sampling rates up to 8~kHz per channel. This high acquisition bandwidth allows direct capture of concurrent high-frequency stimulation waveforms, including temporal interference carrier signals, which is rarely achievable in existing integrated platforms. Experimental results demonstrate accurate sensing, precise stimulation output, and robust concurrent EEG--tES recording without saturation. By combining high performance, compact size, and low system complexity, the proposed platform provides a practical hardware foundation for next-generation closed-loop neuromodulation research and real-world translational applications.

\section{Methods} \label{sec:methods}

\subsection{System Design}

\subsubsection{System architecture overview}

\begin{figure*}
    \centering
    \includegraphics[width=0.63\textwidth]{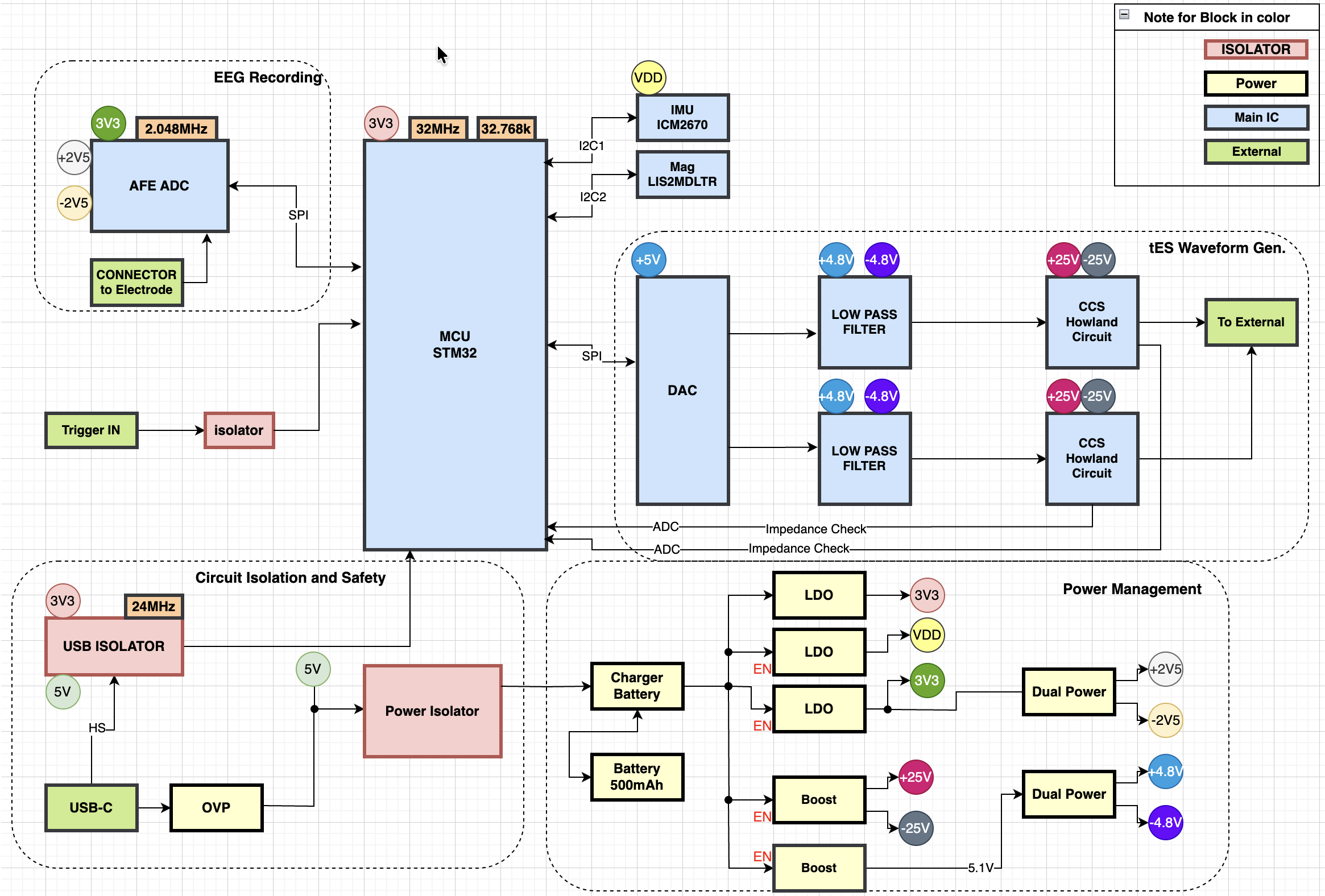}
    \includegraphics[width=0.32\textwidth]{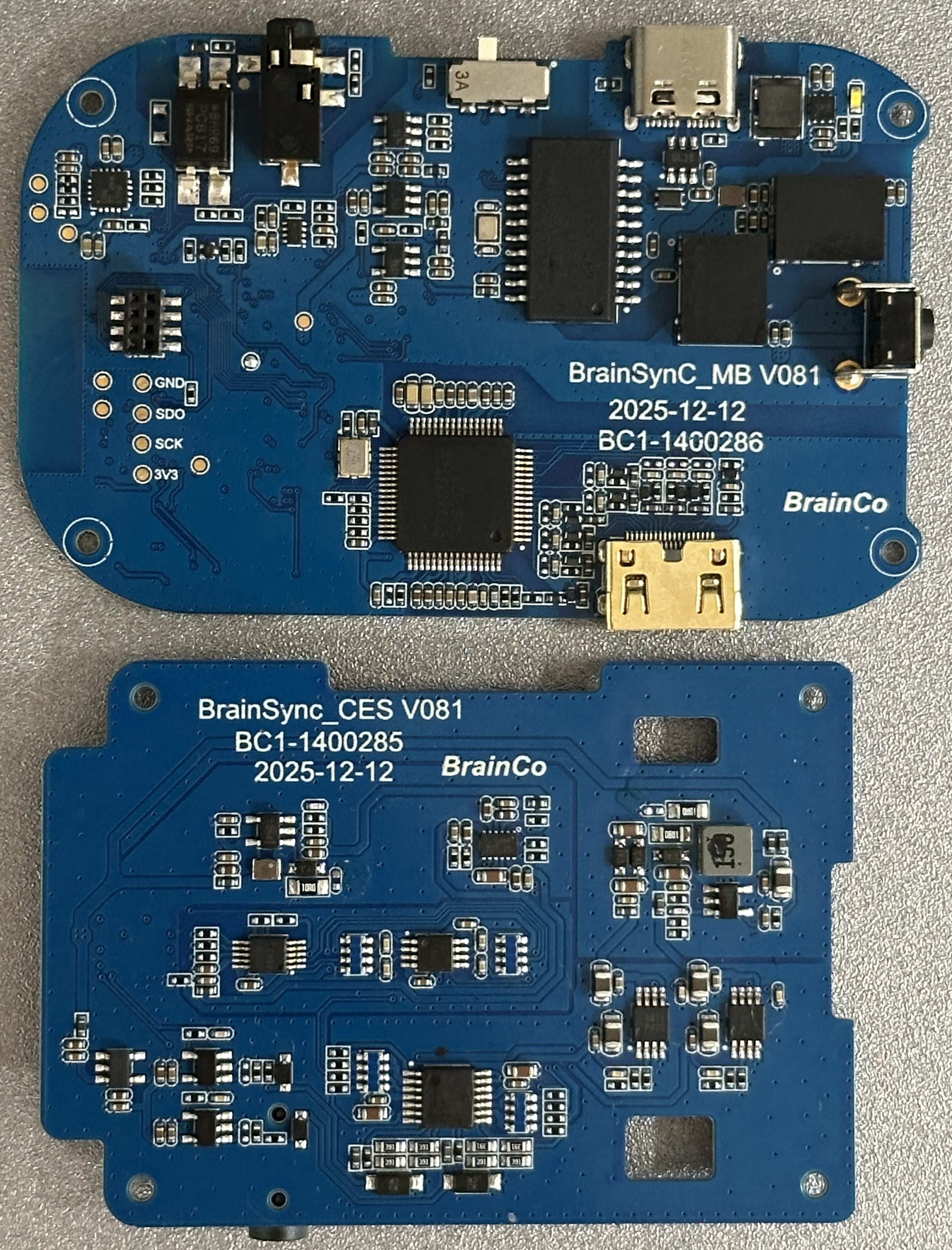}
    \caption{Hardware architecture of the proposed integrated EEG--tES platform (left) and fabricated printed circuit boards (right). The schematic highlights the embedded circuit-level implementation, including EEG acquisition, programmable stimulation generation, galvanic isolation, communication interfaces, and power-management subsystems.}
    \label{fig:system_design}
\end{figure*}

The proposed platform integrates simultaneous electroencephalography (EEG) recording and transcranial electrical stimulation (tES) within a compact embedded system centered on an STM32 microcontroller unit (MCU), as illustrated in \fig{fig:system_design}. The overall architecture is organized into four functional modules: (1) a high-resolution EEG acquisition front-end supporting 8-channel synchronous sampling from 250~Hz to 8~kHz per channel; (2) a programmable tES generation module capable of producing multiple stimulation modalities, including transcranial direct current stimulation (tDCS), transcranial alternating current stimulation (tACS), transcranial temporal interference stimulation (tTIS), ramping profiles, and other customized waveforms; (3) a multilayer galvanic isolation subsystem for participant safety; and (4) a battery-powered dual-rail power management subsystem providing regulated up to $\pm25$~V compliance rails for current stimulation together with low-noise analog supplies for biosignal acquisition. The EEG and stimulation subsystems operate independently under centralized MCU coordination, enabling both open-loop and closed-loop protocols while minimizing mutual interference between sensing and stimulation pathways.

Moreover, a companion host application was developed to provide experiment control, parameter configuration, real-time EEG visualization, and data management. Through the software interface, users can select stimulation modality, current amplitude, waveform frequency, ramping duration, recording mode, and communication interface (USB/Bluetooth). Incoming EEG data are displayed in real time and can be stored for offline analysis. This host-assisted software layer improves usability while preserving low-latency sensing and stimulation control on the embedded device.

\subsubsection{High-speed EEG and concurrent tES waveform acquisition}

The system is designed to acquire neural activity and concurrent high-frequency stimulation waveforms simultaneously, thereby supporting advanced closed-loop neuromodulation studies. The recording front-end is based on the Texas Instruments ADS1299, a 24-bit low-noise delta-sigma analog-to-digital converter optimized for biopotential sensing. All eight differential channels are sampled synchronously at configurable rates from 250~Hz to 8~kHz, substantially exceeding the bandwidth of conventional EEG systems (typically $\leq 1$~kHz). Moreover, our proposed system can achieve $\pm375 mV$ dynamic range to avoid the amplifier saturation during stimulation.

This elevated sampling rate enables simultaneous capture of physiological EEG rhythms and kilohertz-range stimulation carriers used in tTIS. Consequently, high-frequency stimulation artifacts, carrier signals, and their low-frequency interference envelopes can be directly observed in the raw recordings, facilitating artifact modeling, stimulation verification, and the development of future real-time adaptive control algorithms—areas that have been rarely investigated in the existing literature.

To preserve signal quality under wide-band acquisition conditions, each channel incorporates a driven-right-leg (DRL) feedback circuit and configurable analog low-pass filtering. The analog front-end and sampling configuration were designed to preserve both conventional EEG activity and higher-frequency stimulation-related components while attenuating out-of-band noise. Electrode contact quality is continuously monitored using the ADS1299 integrated lead-off detection function, enabling impedance assessment during both recording and stimulation.

Sustained high-rate acquisition is achieved through a co-designed hardware--software streaming pipeline. On the embedded side, the STM32 MCU employs DMA-based data transfer and non-blocking ring-buffer management to maintain continuous 8~kHz multichannel sampling. For high-bandwidth applications, acquired data are streamed through USB, enabling stable transmission of full-rate recordings and concurrent stimulation signals with minimal computational overhead. In addition, the system also supports Bluetooth communication for low-sampling-rate and mobile applications, where wireless operation and user mobility are prioritized over maximum data throughput. On the host side, a companion software platform buffers incoming packets and distributes them to parallel modules for real-time visualization, storage, stimulation control, and optional online processing. Together, this architecture provides stable, low-latency acquisition of both EEG and concurrent tES waveforms, establishing a practical sensing platform for next-generation closed-loop neuromodulation research.

\subsubsection{High-precision programmable tES waveform generation}


The stimulation subsystem generates current-controlled waveforms using a dual-channel 16-bit digital-to-analog converter driven by the STM32 MCU through an SPI interface. The DAC outputs are converted to constant-current stimulation through precision Howland current source circuits capable of delivering up to $\pm4$~mA into loads up to $5~k\Omega$. Our DAC and system were designed to update the DDS waveform at 200 kS/s, enabling fast electrical stimulation waveform generation in real time. According to the transcranial electrical stimulation guidelines in \cite{antal2017low}, the recommended tES current should be kept below 2 mA for safety considerations. Nevertheless, our system is designed to deliver up to 4 mA by default to support more aggressive research studies.

The programmable output stage supports arbitrary waveform synthesis and commonly used tES paradigms, including:

\begin{enumerate}
    \item \textbf{tDCS:} constant-current direct stimulation;
    \item \textbf{tACS:} sinusoidal stimulation with programmable amplitude, frequency, and phase;
    \item \textbf{tTIS:} two independent high-frequency carrier currents (e.g., 2.00~kHz and 2.01~kHz) whose interference produces a low-frequency modulation envelope;
    \item \textbf{Ramping:} programmable soft-start and soft-stop transitions.
\end{enumerate}

To achieve flexible high-fidelity waveform generation without dedicated synthesizer chips, the system adopts an enhanced direct digital synthesis (DDS) architecture implemented entirely on the MCU. DDS is a well-established digital waveform generation technique widely used for precise frequency and phase control \cite{vankka2001direct}. Rather than relying on a static lookup table, two alternating waveform buffers are maintained in memory. While one buffer is streamed to the DAC via DMA at the hardware update rate, the MCU simultaneously computes the next waveform segment in the background. Once transmission of the active buffer is completed, DMA immediately switches to the updated buffer, ensuring continuous and interruption-free stimulation output.

This dual-buffer DDS scheme provides two key advantages. First, it enables real-time generation of adaptive or non-periodic waveforms, including smooth ramping, phase-shifted multichannel tACS, burst modulation, and dynamically adjustable tTIS carriers. Second, it preserves waveform fidelity across a wide frequency range because output generation is not constrained by a fixed lookup-table resolution. Both low-frequency stimulation and kilohertz-range carrier signals can therefore be generated without sacrificing update rate or precision.

Efficient numerical implementations, including lookup-assisted trigonometric computation, ensure that waveform synthesis remains ahead of the DAC output stream under all operating conditions. These hardware--software design choices collectively provide a compact, low-cost, and high-performance platform for programmable tES generation in advanced neuromodulation applications.

\subsubsection{Galvanic isolation and safety considerations}

Participant safety is ensured through comprehensive galvanic isolation between the stimulation/recording electronics and the host interface. All digital communication lines crossing the isolation boundary, including SPI, I$^2$C, and USB, employ high-speed digital isolators. Power domains are independently separated using isolated DC--DC converters and dedicated low-dropout regulators (LDOs) for each subsystem.

The complete analog front-end, including EEG electrodes and stimulation outputs, is electrically floated with respect to earth ground to minimize leakage current pathways. In addition, hardware current limiting, over-voltage protection (OVP), watchdog supervision, and fault recovery mechanisms are implemented to provide fail-safe operation consistent with IEC 60601-2-10 safety principles for medical electrical stimulators. Finally, this power architecture further enhances safety and noise immunity by separating stimulation power delivery from low-voltage digital logic supplies.

\subsection{System Validation}

\subsubsection{EEG Data Acquisition}
To evaluate the signal acquisition accuracy of the proposed EEG device, a cable-based validation experiment was performed under controlled conditions. The device was directly connected to a signal generator (SKX-8000, Ming Sheng Electronic Technology Co.,Ltd, China) via shielded leads. A single-channel pre-recorded EEG waveform was injected as the reference signal, and the device recorded the signal from its input, thereby eliminating physiological and electrode–skin interface variability.

The reference EEG was obtained from a publicly available dataset \cite{Casson2020} used in \cite{kohli2019removal}. The data were recorded from a healthy subject performing sequential eyes-open (60 s) and eyes-closed (30 s) resting-state tasks, totaling approximately 90 s. The eyes-closed segment exhibited prominent alpha-band (around 10Hz) bursts, providing a distinct physiological feature for fidelity assessment.

After recording, both the reference and recorded signals were processed identically. A notch filter was applied to suppress powerline interference, followed by z-score normalization to remove amplitude scaling differences and DC offsets. Signal similarity was quantified using the Pearson correlation coefficient (CC), and waveform preservation was visually verified in the time domain. This controlled signal injection approach enables quantitative assessment of sensing accuracy while maintaining physiologically meaningful signal characteristics.

\subsubsection{tES waveform accuracy via oscilloscope recording}

To validate the transcranial electrical stimulation (tES) output performance, the device was connected to an oscilloscope (Tektronix MSO44B, 1GHz, 6.25GS/s, USA) through shielded cables. All stimulation waveforms were generated using our self-developed control software, which allows programmable adjustment of stimulation type, current amplitude, frequency, and ramping profile. A fixed resistive load of $5K\Omega$ was used during all measurements to emulate typical electrode–skin impedance encountered in practical tES applications. This controlled load condition ensures repeatable and comparable electrical characterization.

The output accuracy of the proposed stimulator was evaluated using a resistive load measurement setup. A precision resistor load of 5~k$\Omega$ was connected to the stimulator output, and the voltage waveform across the load was recorded using a digital oscilloscope. The output current was calculated from the measured voltage according to Ohm's law.

For tDCS, four current amplitudes (0.5, 1, 2, and 4~mA) were tested. 

For tACS, four current amplitudes (0.5, 1, 2, and 4~mA) were evaluated at four stimulation frequencies (5, 10, 20, and 40~Hz), resulting in 16 test conditions.

For tTIS, two independent current sources were used to generate the interference waveform. 
Four current amplitudes (0.5, 1, 2, and 4~mA) were tested with carrier frequency pairs designed to produce envelope frequencies of 5, 10, 20, and 40~Hz, corresponding to frequency pairs of 
2000/2005~Hz, 2000/2010~Hz, 2000/2020~Hz, and 2000/2040~Hz, respectively.

The generated waveforms were captured by the oscilloscope and exported for post-processing in MATLAB (The MathWorks Inc, USA). For each test condition, the following parameters were quantified: output current amplitude, amplitude error relative to the programmed value, frequency spectrum, and spectral error. For the current output accuracy quantitation, the current amplitude error and the current frequency error were calculated, based on the following equations, similar to the work in \cite{zhang2022designing}:

\begin{equation}
I_{\mathrm{err},\%}= \frac{\left| I_{\mathrm{amp,measured}} - I_{\mathrm{amp,set}} \right|}
{I_{\mathrm{amp,set}}}
\times 100\%
\end{equation}

\begin{equation}
f_{\mathrm{err},\%}=\frac{\left| f_{\mathrm{measured}} - f_{\mathrm{set}} \right|}
{f_{\mathrm{set}}}
\times 100\%
\end{equation}

\subsection{On Gelatine Head Phantom Recording Performance Evaluation}
To validate the device’s sensing performance under realistic biological conditions, a gelatine-based head phantom was employed to emulate the electrode–tissue interface. This setup enabled evaluation of both EEG-only acquisition and concurrent EEG–tES recording performance. The primary objectives were to (1) assess EEG signal fidelity under biologically representative impedance conditions and (2) demonstrate stable recording of concurrent EEG and high-amplitude tES signals without amplifier saturation, thereby verifying the effectiveness of the high sampling rate and wide dynamic input range of the system.

\begin{figure}
    \centering
    \includegraphics[width=0.49\textwidth]{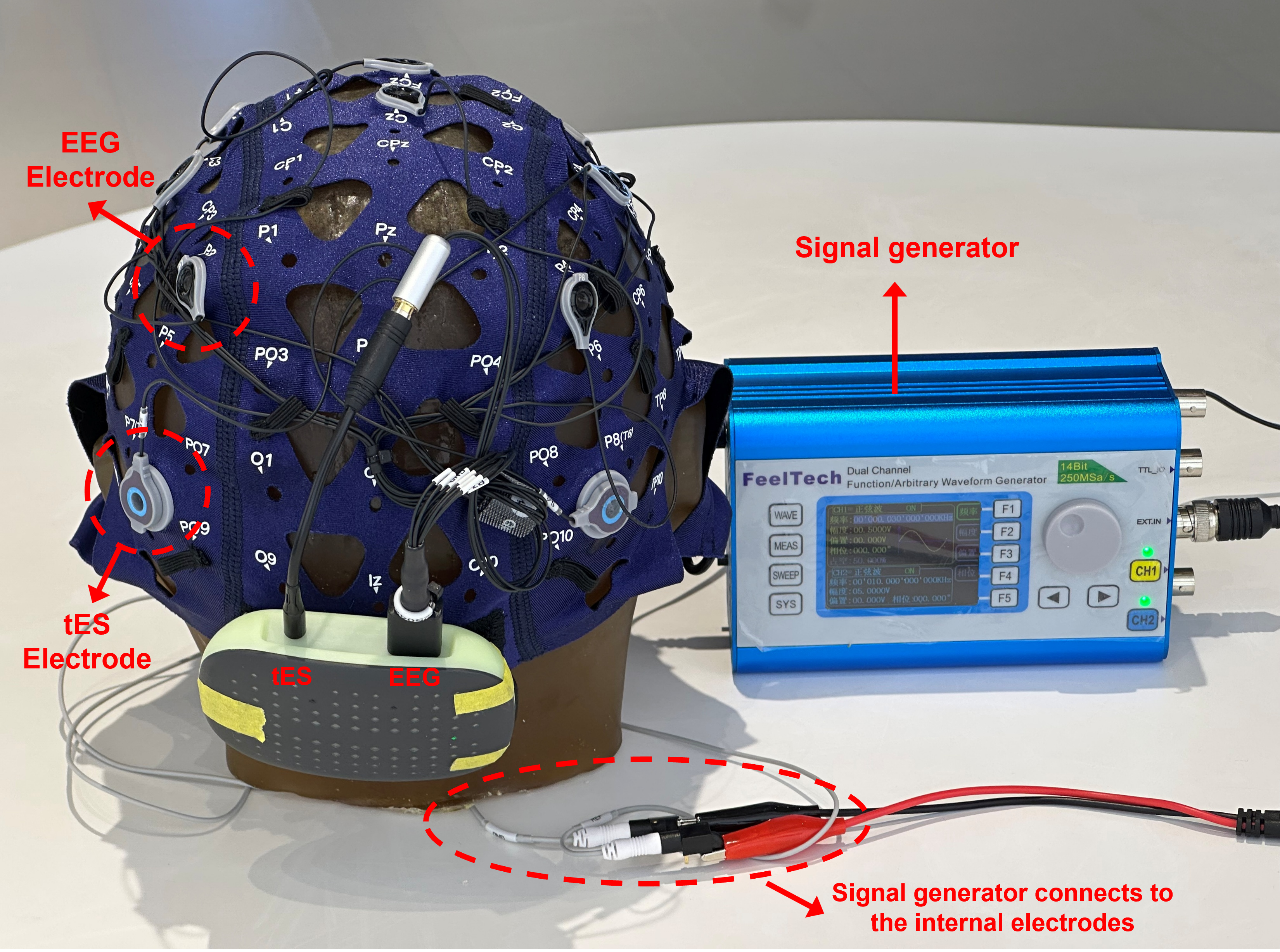}
    \caption{Gelatine head phantom setup for simulated EEG and concurrent EEG-tES recording. The assembled prototype device measures around 87 mm × 63 mm × 26 mm.}
    \label{fig:phantom_setup}
\end{figure}

The whole head phantom setup can be seen from \fig{fig:phantom_setup}. The head phantom was fabricated using gelatine powder, water, and NaCl, following the formulation reported in \cite{owda2021investigating}. During fabrication, two standard Ag/AgCl cup electrodes were embedded within the phantom volume. These electrodes were connected to the other signal generator (FeelTech, FY6300P, China), which replayed the same pre-recorded EEG waveform described in the previous section, thereby serving as a controlled internal signal source.

An EEG cap was positioned on the phantom surface, and the proposed EEG–tES device was mounted on the cap. Eight Ag/AgCl EEG electrodes (Greentek Pty Ltd. Wuhan, China) were symmetrically placed according to the international 10–10 system (locations at: Fp1, Fp2, F3, F4, C3, C4, P3 and P4). tES electrodes (Ag/AgCl electrodes with rubber holders and enlarged contact areas from Greentek Pty Ltd.) were positioned at locations F7, P7, F8, P8, to construct 2 electrical channels: channel 1: F7 and P7, channel 2: F8 ad P8. Ag/AgCl electrodes with conductive gel were adopted to reduce the electrode contact impedance \cite{xing2021opportunities}. The lower interface impedance helps ensure that the delivered stimulation current more closely matches the programmed output during operation . This system validation setup was widely used for bioelectronic system validation \cite{xing20233d, kohli2019removal}. In this paper, we validated our system via both EEG recording and concurrent EEG+tES recording on the head phantom.

\subsubsection{EEG recording on phantom}
The EEG signals were acquired at a sampling rate of 8000 Hz across eight channels. Following acquisition, both the recorded signal and the input template signal were resampled to 200 Hz to ensure a common temporal resolution. Each signal was then band-pass filtered between 1–40 Hz using a 9th-order elliptic IIR filter applied in zero-phase mode to avoid phase distortion. Subsequently, both signals were standardized using z-score normalization to mitigate differences in amplitude scaling. Pearson correlation was computed using a sliding-window approach, in which the template signal was shifted across the recorded signal to identify the segment with maximal overlap, and the corresponding correlation coefficient was reported.

\subsubsection{Simultaneous EEG and tES recording on phantom}
To validate the device’s capability for concurrent EEG recording during various tES waveforms, experiments were conducted using the gelatine-based head phantom described in the previous section. The simulated EEG signal was generated through the embedded internal electrode, as described above, while transcranial electrical stimulation (tES) was delivered using our self-developed control software. The stimulation waveforms were configured sequentially as follows: tDCS (1 mA), tACS (1 mA, 20 Hz), and temporally interfering stimulation (tTIS, 1 mA, 2000/2040 Hz). Note that tDCS and tACS only used 1 stimulation channel: F7-P7 pair, whereas tTIS used 2 stimulation channel pairs. During stimulation, the EEG amplifier simultaneously recorded both the EEG signal and the stimulation-induced artifacts under high sampling-rate settings.

The primary objective of this experiment was to verify stable data acquisition under concurrent EEG–tES conditions, particularly in the presence of high-amplitude and high-frequency stimulation artifacts. The evaluation focused on confirming that the amplifier operated without saturation, clipping, or data packet loss, while maintaining sufficient dynamic range to capture broadband signals. After acquisition, the raw signals were directly plotted without additional preprocessing, and fast Fourier transform (FFT) analysis was performed to inspect the spectral components of the recorded concurrent EEG–tES signals. This experiment demonstrates the system’s ability to reliably record broadband, high-dynamic-range signals during active stimulation, which remains a major challenge for many existing EEG–tES systems.

\subsection{Computational performance evaluation}
The computational performance of the proposed system was characterized using four metrics: latency, CPU usage, power consumption, and functional synchronization.

\begin{itemize}
    \item \textbf{Communication latency:} Defined as the elapsed time between the user command to initiate data streaming in the host software and reception of the first valid EEG data packet at the host computer. 

    \item \textbf{CPU usage:} Estimated in firmware from the proportion of processor usage during steady-state operation. The CPU usage values were periodically reported through the embedded diagnostic logging interface.

    \item \textbf{Power consumption:} Evaluated during battery-powered operation by measuring the discharge current drawn from the 3.7~V lithium battery under representative operating conditions and converting the measured current to power consumption, reported in milliwatts (mW).

    \item \textbf{Synchronization delay:} Defined as the temporal offset between the onset of EEG sampling and the onset of stimulation output when both subsystems were triggered simultaneously. This metric was measured internally using firmware timestamps to assess coordination accuracy between sensing and stimulation functions.
\end{itemize}

\section{Results} \label{sec:results}

\subsection{EEG hardware validation} \label{eeghardware}

\begin{figure*}
    \centering
    \includegraphics[width=0.8\textwidth]{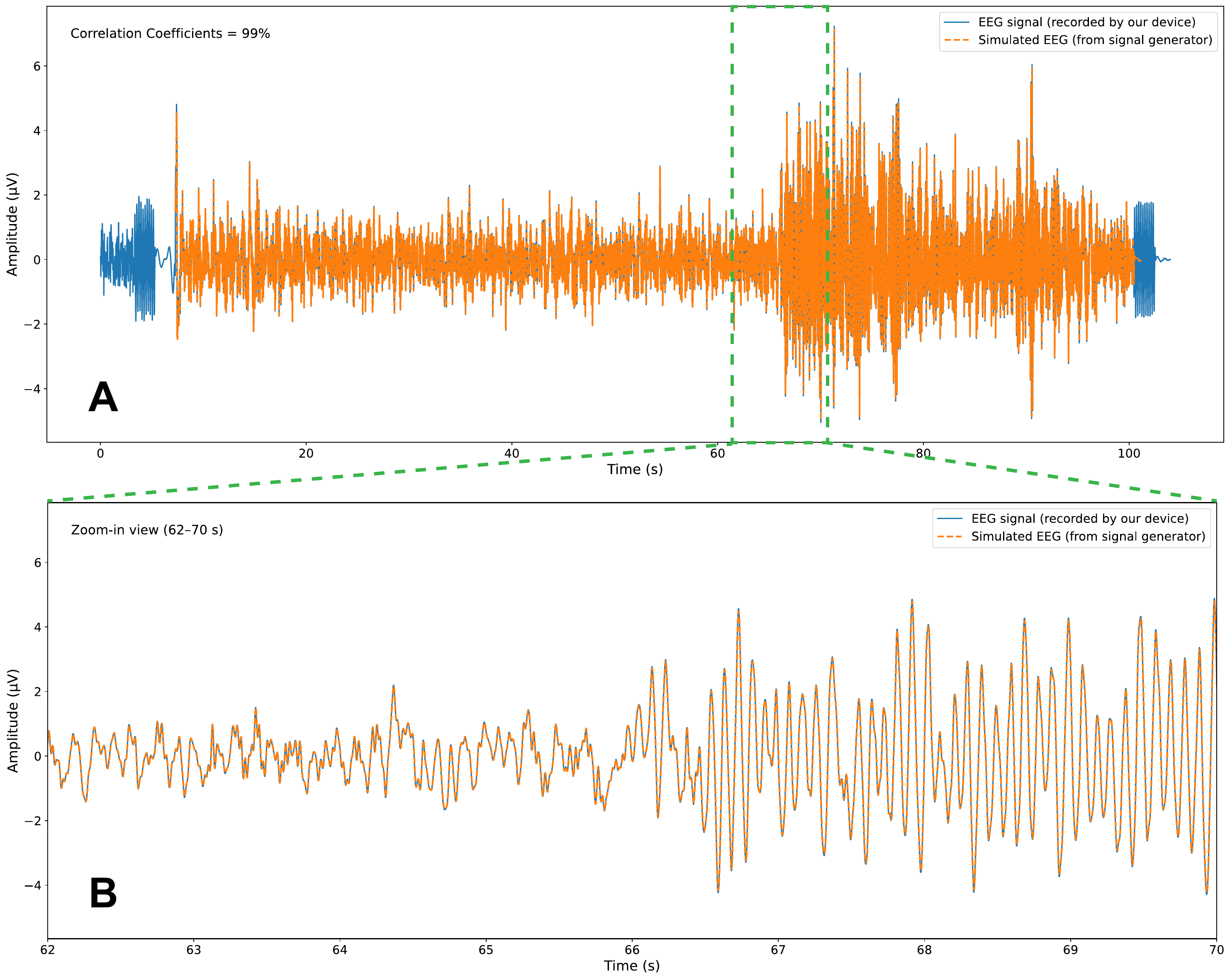}
    \caption{Direct cable-based EEG validation using a signal generator. (A) Comparison between the EEG waveform recorded by the proposed device (blue solid line) and the reference input waveform (orange dashed line). Distinct alignment markers were added at the beginning and end of the signal for synchronization. (B) Enlarged view demonstrating the high similarity between the recorded and reference signals.}
    \label{fig:EEG_signal_generator}
\end{figure*}

The baseline sensing performance of the proposed EEG acquisition subsystem was first evaluated under controlled electrical conditions. As shown in \fig{fig:EEG_signal_generator}, a reference EEG waveform generated by the signal generator was directly connected to the device input through a cable connection, thereby eliminating variability introduced by electrode contact impedance, biological tissue conduction, and environmental motion artifacts.

Under this ideal test condition, the recorded signal closely matched the injected EEG template in both waveform morphology and temporal alignment. The time-domain Pearson correlation coefficient reached $99\%$, demonstrating highly accurate signal acquisition by the proposed system. In addition to the strong correlation, the recorded waveform preserved the amplitude trend and phase characteristics of the input signal, with no observable distortion, clipping, or timing offset.

These results verify the intrinsic sensing fidelity of the analog front-end, analog-to-digital conversion stage, and embedded data transmission pipeline. They further indicate that the system can faithfully acquire broadband EEG signals. This controlled benchmark provides a reliable reference for the subsequent phantom-based and simultaneous EEG--tES experiments, where larger stimulation artifacts, volume-conduction effects, and wider dynamic-range requirements are expected.

\subsection{tES Waveform Fidelity Validation}

\begin{table*}[htbp]
\label{tab:stim_output_accuracy}
\centering
\caption{Measured output current amplitude error (\%) for tDCS, tACS, and tTIS under a 5~k$\Omega$ resistive load. For tTIS, the reported values represent the mean error of the two stimulation channels. 
Frequency error for tACS and tTIS was negligible in all tested conditions and is therefore described in the text rather than tabulated.}
\label{tab:stim_output_accuracy}

\renewcommand{\arraystretch}{1.2}
\setlength{\tabcolsep}{8pt}

\textbf{(a) tDCS}

\begin{tabular}{cc}
\hline
Set current (mA) & Amplitude error (\%) \\
\hline
0.5 & 0.06 \\
1   & 0.34 \\
2   & 0.12 \\
4   & 0.37 \\
\hline
\end{tabular}

\vspace{1em}

\textbf{(b) tACS}

\begin{tabular}{ccccc}
\hline
Set current (mA) & \multicolumn{4}{c}{Amplitude error (\%)} \\
\cline{2-5}
 & 5 Hz & 10 Hz & 20 Hz & 40 Hz \\
\hline
0.5 & 0.38 & 0.44 & 0.38 & 0.36 \\
1   & 0.47 & 0.48 & 0.48 & 0.54 \\
2   & 0.27 & 0.27 & 0.25 & 0.29 \\
4   & 0.41 & 0.41 & 0.43 & 0.44 \\
\hline
\end{tabular}

\vspace{1em}

\textbf{(c) tTIS *}

\begin{tabular}{ccccc}
\hline
Envelope frequency (Hz) & 5 Hz & 10 Hz & 20 Hz & 40 Hz \\
Carrier pair (Hz) & 2000/2005 & 2000/2010 & 2000/2020 & 2000/2040 \\
\hline
Set current (mA) & \multicolumn{4}{c}{Amplitude error (\%)} \\
\hline
0.5 & 0.61 & 0.56 & 0.53 & 0.86 \\
1   & 0.72 & 0.43 & 0.42 & 0.33 \\
2   & 0.86 & 0.46 & 0.42 & 0.41 \\
4   & 0.61 & 0.53 & 0.51 & 0.58 \\
\hline
\end{tabular}

\scriptsize{(* For tTIS, the reported values represent the maximum absolute error for each of the two stimulation channels.)}
\end{table*}

\subsubsection{tDCS}
\begin{figure}
    \centering
    \includegraphics[width=0.49\textwidth]{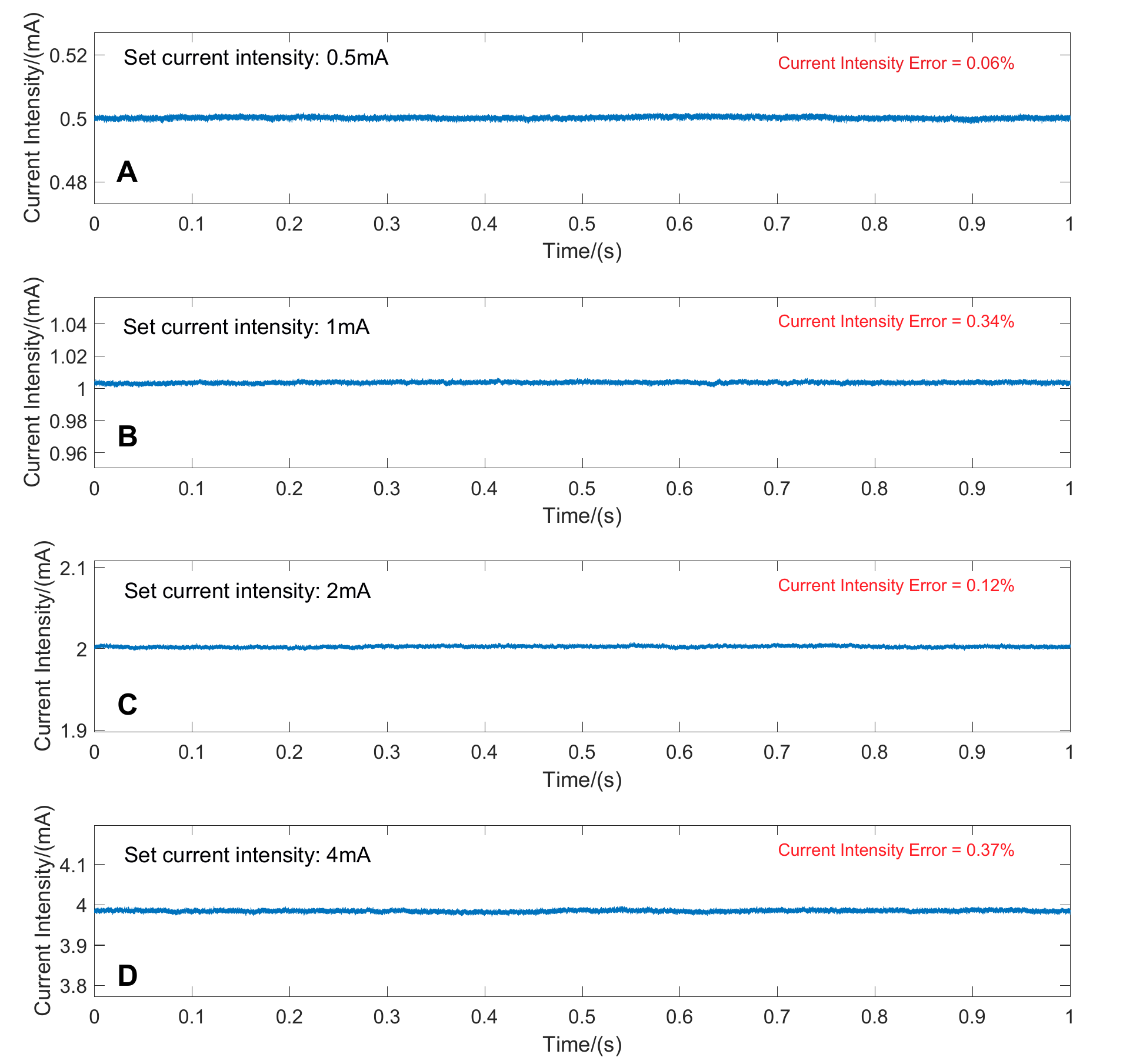}
    \caption{tDCS waveform validation on osciliscope, with different current intensities: (A) 0.5mA; (B) 1mA; (C) 2mA; (D) 4mA.}
    \label{fig:firmware_tDCS}
\end{figure}

The direct current stimulation output was first evaluated under resistive load conditions. As shown in \fig{fig:firmware_tDCS}, representative tDCS waveforms were measured at programmed current intensities of 0.5, 1, 2, and 4~mA using the oscilloscope. Across all tested settings, the system produced stable constant-current outputs with minimal fluctuation over time.

Quantitative analysis showed that the output current amplitude error remained below $1\%$ for all tested conditions (seen in \tab{tab:stim_output_accuracy}). These results demonstrate accurate current regulation across the full operating range and confirm the suitability of the stimulation module for applications requiring precise tDCS delivery.

\subsubsection{tACS}
\begin{figure*}
    \centering
    \includegraphics[width=\textwidth]{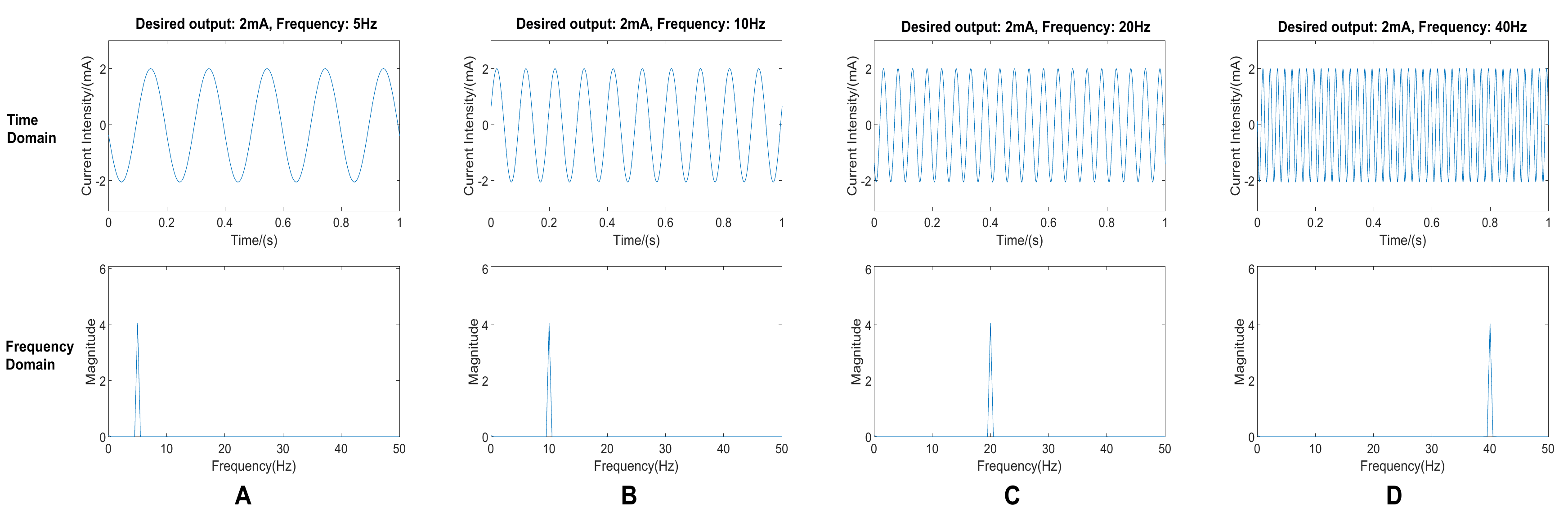}
    \caption{tACS waveforms validation with fixed current intensity but different frequency outputs. (A) 2mA, 5Hz; (B) 2mA, 10Hz; (C) 2mA, 20Hz; (D) 2mA, 40Hz; (Upper row: time domain tACS waveforms; Lower row: the FFT features in the frequency domain for each frequency outputs.)}
    \label{fig:firmware_tACS}
\end{figure*}

Representative tACS outputs at a fixed current intensity of 2~mA and frequencies of 5, 10, 20, and 40~Hz are shown in \fig{fig:firmware_tACS}. Time-domain measurements indicate that the generated waveforms closely follow the intended sinusoidal profiles across all tested frequencies.

The measured current amplitude error remained below $1\%$ for all tested conditions. A comprehensive evaluation across multiple current--frequency combinations is summarized in \tab{tab:stim_output_accuracy}. Frequency-domain analysis (FFT), shown in the lower row of \fig{fig:firmware_tACS}, reveals sharp spectral peaks aligned with the programmed stimulation frequencies, with no observable frequency drift and minimal harmonic distortion.

These results confirm that the proposed stimulator maintains both amplitude accuracy and spectral fidelity during oscillatory stimulation, which is essential for rhythm-specific neuromodulation studies. More test results are provided in Supplementary Materials.

\subsubsection{tTIS}
\begin{figure*}
    \centering
    \includegraphics[width=\textwidth]{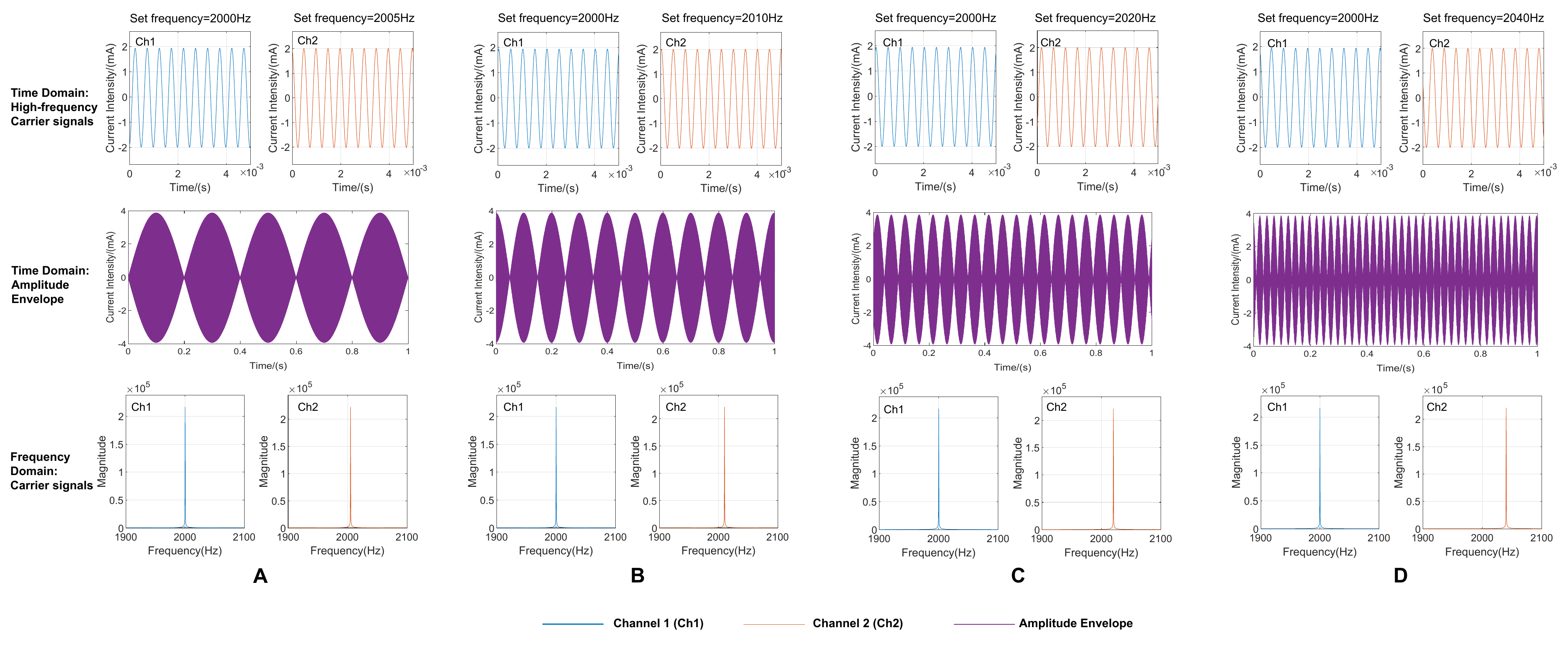}
    \caption{tTIS waveform validation at a fixed current intensity of 2~mA using different carrier-frequency pairs: (A) 2000/2005~Hz; (B) 2000/2010~Hz; (C) 2000/2020~Hz; (D) 2000/2040~Hz. Top row: channel 1 and channel 2 carrier waveforms. Middle row: resulting amplitude envelope by linearly adding 2 tTIS channel outputs. Bottom row: FFT spectra of the two carrier outputs.}
    \label{fig:firmware_tTIS}
\end{figure*}

Representative temporal interference stimulation (tTIS) outputs at a current intensity of 2~mA are presented in \fig{fig:firmware_tTIS}. Two independent stimulation channels generate sinusoidal carrier currents in the kilohertz range (e.g., 2000~Hz and 2005--2040~Hz), as shown in the top row. Across all tested carrier-frequency combinations, the measured current amplitude error for both channels remained below $1\%$, indicating accurate high-frequency current generation. When the two carrier signals are spatially combined, a low-frequency amplitude-modulated envelope is formed, as illustrated in the middle row of \fig{fig:firmware_tTIS}. The resulting envelope exhibits stable periodic modulation consistent with the theoretical temporal interference principle. Complete quantitative results for all tested conditions are summarized in \tab{tab:stim_output_accuracy}.

Frequency-domain analysis shown in the bottom row demonstrates distinct spectral peaks at the programmed carrier frequencies, with no observable frequency deviation. These findings verify precise control of both carrier outputs and demonstrate the capability of the proposed system to generate high-fidelity tTIS waveforms, which are not supported by many conventional integrated EEG--tES platforms. More test results are provided in Supplementary Materials.

\subsubsection{Ramping}
\begin{figure}
    \centering
    \includegraphics[width=0.49\textwidth]{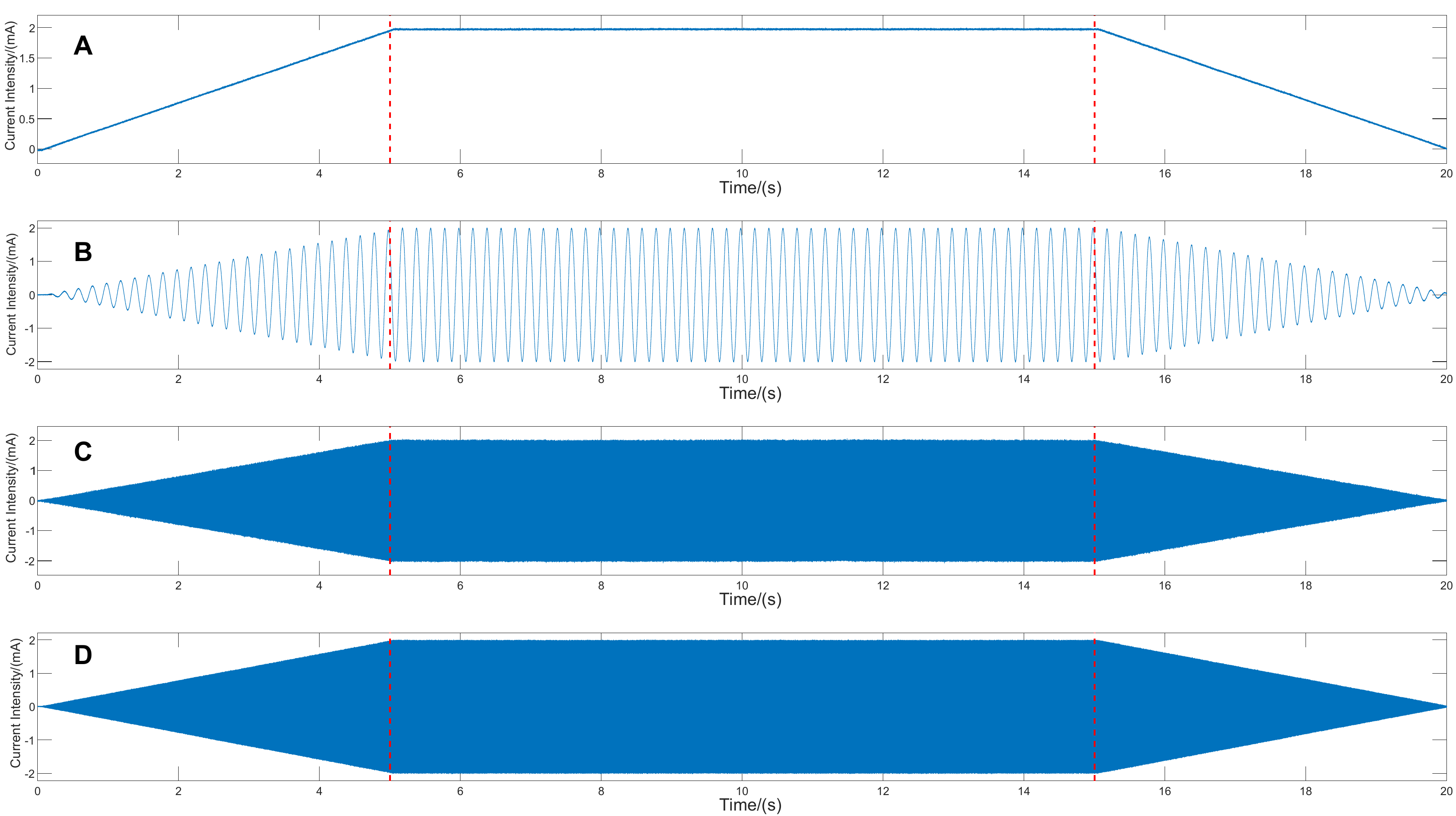}
    \caption{Programmable 5~s ramp-up and 5~s ramp-down profiles for different stimulation modalities. (A) tDCS: 2~mA; (B) tACS: 2~mA; (C) 2~kHz tTIS carrier; (D) 2.01~kHz tTIS carrier. Ramp duration can be adjusted according to experimental requirements.}
    \label{fig:firmware_ramping}
\end{figure}

The programmable ramping function was evaluated for tDCS, tACS, and both tTIS output channels, with representative examples shown in \fig{fig:firmware_ramping}. In this experiment, the stimulation current gradually increased from zero to the target amplitude during a 5~s ramp-up phase and symmetrically decreased during a 5~s ramp-down phase.

The measured waveforms show smooth and monotonic transitions without discontinuities, overshoot, or instability, indicating reliable real-time control of output amplitude. Such gradual onset and offset profiles are important for reducing abrupt sensory perception, improving user comfort, and minimizing transient artifacts during stimulation experiments.

\subsection{On-Phantom Signal Recording}

\subsubsection{EEG recording}

\begin{figure}
    \centering
    \includegraphics[width=0.49\textwidth]{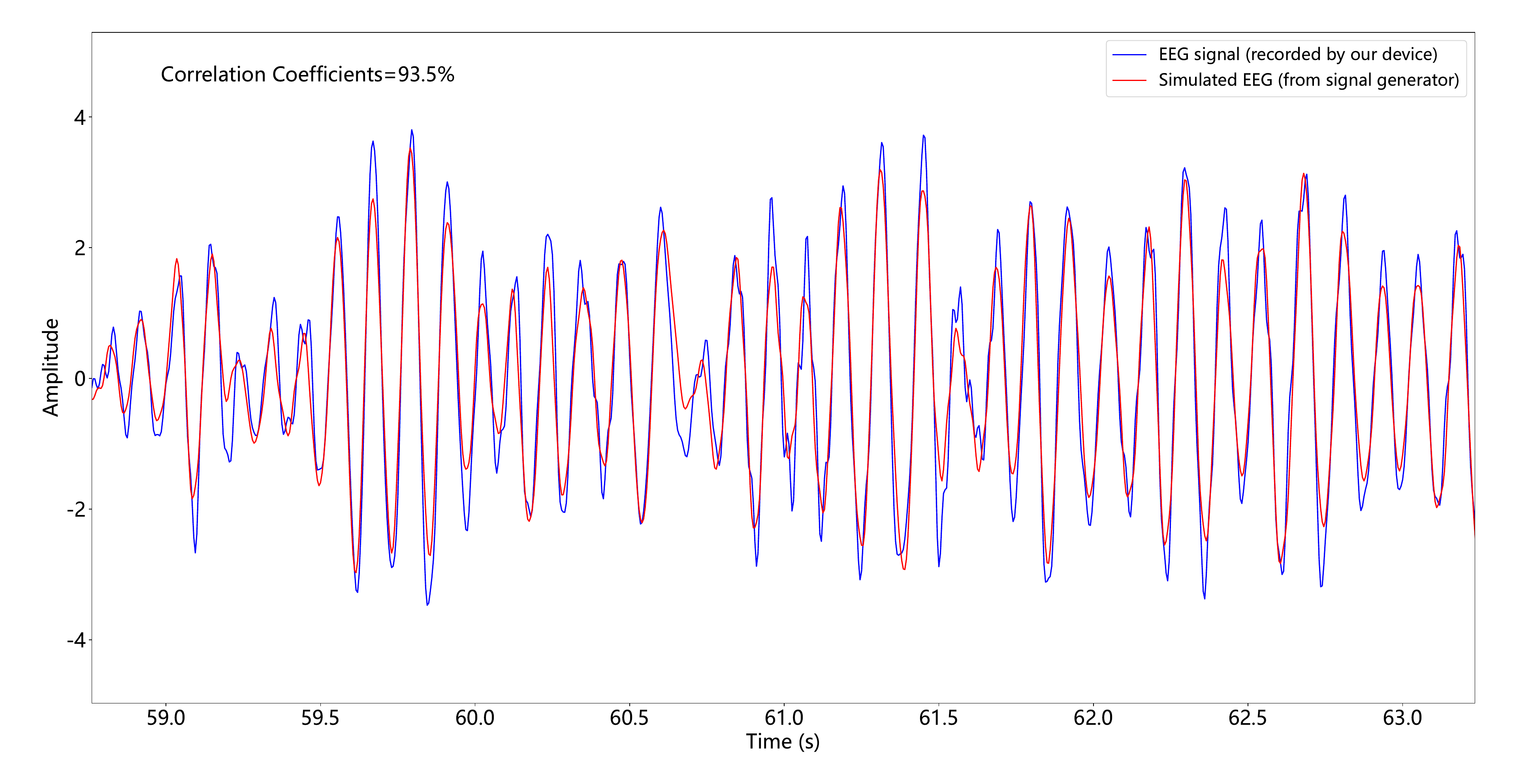}
    \caption{EEG-only recording using the gelatine head phantom. The recorded signal shows high similarity to the injected reference waveform, with a correlation coefficient of $93.5\%$.}
    \label{fig:phantom_EEG}
\end{figure}

The sensing performance of the proposed system was further evaluated using the gelatine head phantom under biologically representative conditions. A representative segment of the recorded EEG signal is shown in \fig{fig:phantom_EEG}. The Pearson correlation coefficient between the injected reference waveform and the recorded EEG signal reached $93.5\%$, indicating strong signal fidelity after transmission through the phantom medium.

For comparison, the direct cable-based validation described in \sect{eeghardware} achieved a higher correlation coefficient of $99\%$ under ideal electrical conditions. The moderate reduction observed in the phantom experiment is expected and primarily reflects the additional complexities introduced by the conductive medium. Specifically, the gelatine phantom introduces volume-conduction effects, electrode--interface impedance, environmental interference, and distributed resistive--capacitive characteristics, all of which can attenuate or slightly distort the transmitted waveform.

Despite these practical effects, the recorded signal remained highly consistent with the reference input, demonstrating robust acquisition performance under more realistic measurement conditions. The phantom setup therefore provides a controllable and repeatable intermediate validation platform between benchtop electrical testing and future human experiments. These results further support the suitability of the proposed device for simultaneous EEG--tES studies, where stable sensing performance must be maintained in the presence of complex conductive pathways and large stimulation artifacts.

\subsubsection{Concurrent EEG+tES recording}
\begin{figure*}[!t]
    \centering
    \includegraphics[width=\textwidth]{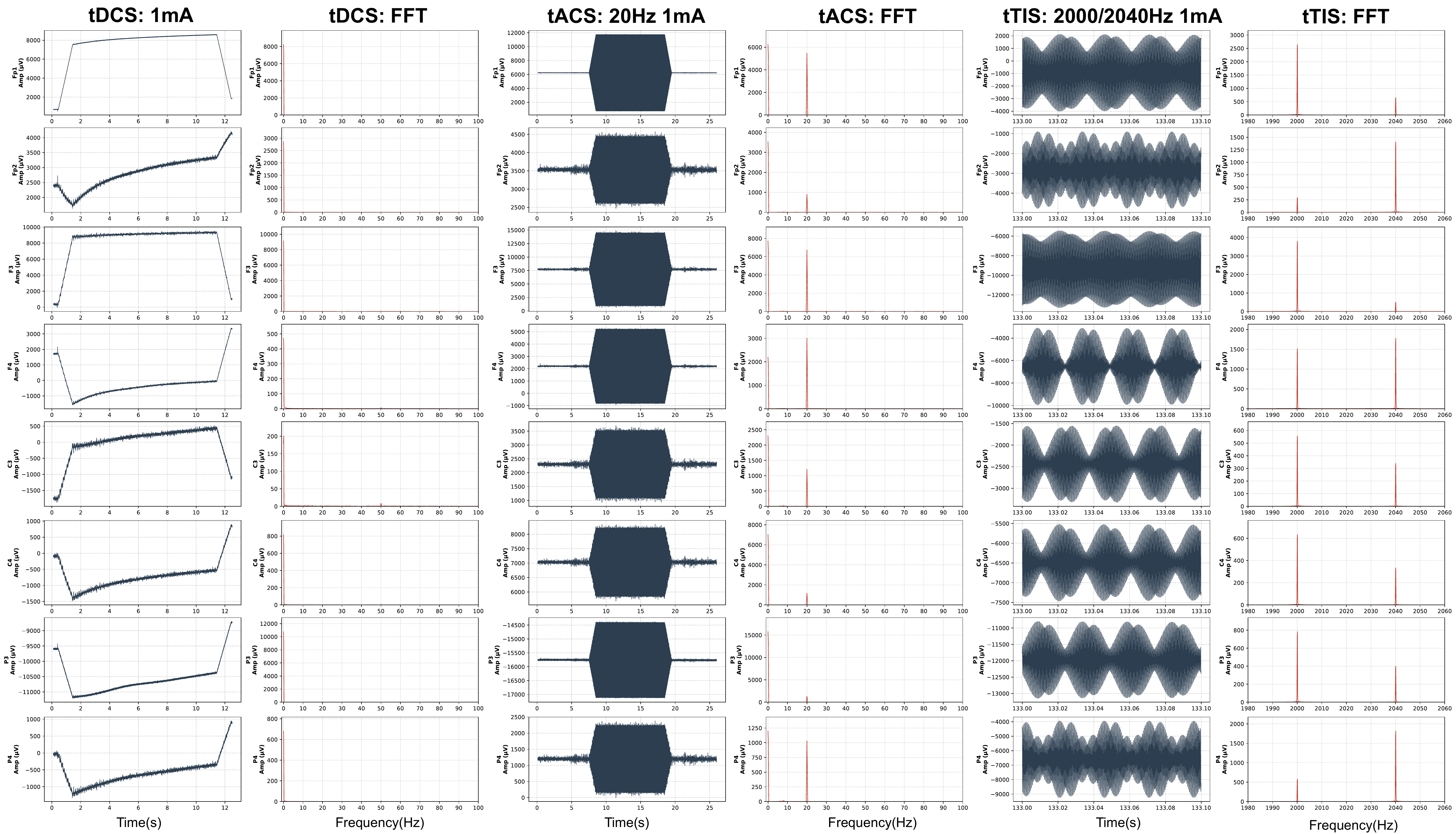}
    \caption{Example of raw simultaneous EEG and tES recording in a gelatine head phantom under 8~KHz sampling rate, without any processing steps (EEG signals are corrupted by large tES artifacts). Column 1-2: Recorded EEG–tDCS mixture at 1 mA and its corresponding frequency spectrum (FFT). Column 3-4: Recorded EEG–tACS mixture at 1 mA, 20 Hz, with the corresponding FFT shown on the right. Column 5-6: Recorded EEG–tTIS mixture at 1 mA with carrier frequencies of 2000 Hz and 2040 Hz, and the corresponding FFT.}
    \label{fig:eeg_tes}
\end{figure*}

The simultaneous EEG and transcranial electrical stimulation (tES) recording performance was evaluated, with representative results shown in \fig{fig:eeg_tes}. All signals presented are raw recordings without post-processing. Because the amplifier operates in a DC-coupled configuration, the signals were not baseline-normalized, and soft ramping was applied under all stimulation conditions. The even-numbered columns of \fig{fig:eeg_tes} show the corresponding frequency-domain representations (FFT) of the recorded signals, each exhibiting the expected spectral peaks consistent with the applied stimulation parameters.

Compared with the injected EEG signals generated by the signal generator at microvolt-to-millivolt amplitudes, the applied stimulation signals are substantially larger, reaching the volt range. For example, a 2 mA current delivered across an approximately 5 k$\Omega$ load corresponds to a voltage of about 10 V. As a result, the recorded EEG signals are strongly dominated by concurrent stimulation artifacts.

For tDCS, the constant stimulation current introduces a substantial DC offset, shifting the signal baseline upward or downward depending on the stimulation polarity, as observed across different EEG channels. In the case of tACS, the sinusoidal stimulation waveform largely obscures the underlying EEG activity, with residual EEG features only occasionally visible near the waveform extrema. For tTIS, the recorded signals clearly capture the superposition of two high-frequency carrier signals, producing channel-dependent amplitude envelopes. The FFT results further confirm that these waveforms consist of the expected carrier-frequency components with varying relative intensities.

Across all tES modalities, eight EEG channels were recorded simultaneously. Waveform characteristics differ across channels because of variations in electrode placement, inter-electrode distance, and stimulation polarity. As reported in \cite{kohli2019removal}, electrode montage plays a critical role in concurrent EEG–tES recordings; symmetric electrode configurations are recommended to minimize spatial bias and reduce the risk of amplifier saturation.

Importantly, these results demonstrate that the proposed system can reliably capture mixed EEG–tES waveforms across multiple stimulation modalities under high sampling-rate conditions without signal clipping or loss of critical features, which most current EEG devices cannot achieve. This capability provides a robust experimental foundation for future studies, particularly in the development of closed-loop neuromodulation systems, where accurate real-time acquisition of stimulation-contaminated EEG signals is essential \cite{noury2016physiological}.

\subsection{Computational Performance Evaluation}  

\tab{tab:computational_performance} summarizes the computational performance of the proposed system under representative operating conditions. Across all tested modes, the platform maintained stable operation with moderate processor utilization, low synchronization delay, and consistent startup latency, supporting its suitability for portable multimodal neuromodulation applications.

The measured startup latency ranged from 1.56~s to 1.67~s across BLE and USB communication modes. The relatively small variation indicates robust command execution and reliable data-stream initialization under different transmission interfaces. USB operation showed slightly lower latency than BLE mode, which is expected due to the reduced communication overhead of wired transmission.

CPU utilization remained below 40\% during standard 8-channel EEG acquisition at 500~Hz, and below 53\% even under the most demanding condition of 8-channel EEG acquisition at 8~kHz with concurrent tTIS output. These results indicate sufficient computational headroom for stable long-term operation and future implementation of additional onboard processing or adaptive control algorithms.

During standalone battery-powered BLE operation, measured power consumption was 139.3~mW in EEG-only mode and 249.6~mW during simultaneous EEG acquisition with 2~mA tACS output. The moderate power demand supports the feasibility of wearable and mobile use cases, while the increase during stimulation is consistent with the expected load of the current-output stage.

The synchronization delay between EEG acquisition onset and stimulation onset remained below 1~ms across all tested conditions. Such temporal precision is essential for paradigms requiring tightly coordinated sensing and stimulation, and validates the effectiveness of the unified MCU-based scheduling architecture.

Overall, the proposed system demonstrated reliable communication performance, efficient resource utilization, low-power wireless operation, and sub-millisecond synchronization accuracy, supporting its practical use as an integrated EEG--tES platform for both laboratory and mobile environments.

\begin{table*}[t]
    \centering
    \caption{Computational performance of the proposed system under representative operating conditions. Reported values correspond to approximate steady-state readouts obtained during continuous operation, as latency, CPU load, and power consumption may vary over time during runtime. The selected test modes cover mobile wireless use, standard wired acquisition, high-rate sensing, and concurrent EEG--tES operation.}
    \label{tab:computational_performance}
    \renewcommand{\arraystretch}{1.2}
    \begin{tabular}{p{7.0cm} p{2.0cm} p{2.0cm} p{2.0cm} p{2.0cm}}
    \hline
    \textbf{Condition / System Configuration} & \textbf{Latency (s)} & \textbf{CPU Load (\%)} & \textbf{Power (mW)} & \textbf{Sync Offset (ms)} \\
    \hline
    
    BLE, 8-ch EEG @ 500 Hz & 1.67 & $34\%$ & 139.3 & $<1\,\mathrm{ms}$ \\
    
    BLE, 8-ch EEG @ 500 Hz + tACS (2 mA) & 1.67 & $36\%$ & 249.6 & $<1\,\mathrm{ms}$ \\
    
    USB, 8-ch EEG @ 500 Hz & 1.56 & $38\%$ & -- & $<1\,\mathrm{ms}$ \\
    
    USB, 8-ch EEG @ 8 kHz & 1.58 & $52\%$ & -- & $<1\,\mathrm{ms}$ \\
    
    USB, 8-ch EEG @ 8 kHz + tTIS (2000/2040 Hz, 2 mA) & 1.59 & $53\%$ & -- & $<1\,\mathrm{ms}$ \\
    
    \hline
    \end{tabular}
    
    \vspace{2pt}
    \scriptsize{* Under USB connection mode, the device is powered and charged via USB; therefore, standalone power consumption was not measured.}
\end{table*}

\section{Discussions}
This work demonstrates that high-performance integrated EEG--tES systems can be realized using a compact microcontroller-based architecture without relying on power-intensive FPGA implementations. The experimental results show that the proposed system is capable of simultaneously supporting multichannel EEG acquisition at sampling rates up to 8~kHz together with accurate programmable stimulation output across multiple tES modalities. These findings are significant because they challenge the common assumption that high-rate sensing and flexible waveform generation necessarily require more complex digital hardware.

A key contribution of the proposed design is the efficient use of embedded resources through hardware--software co-optimization. By combining DMA-based data transfer, interrupt-efficient scheduling, ring-buffer streaming, and MCU-based DDS waveform synthesis, the system achieves reliable real-time operation within the computational constraints of a microcontroller. This approach reduces hardware complexity, power consumption, and component count while preserving sufficient throughput for concurrent EEG recording and tES delivery. Compared with FPGA-based implementations, the proposed architecture provides practical advantages in portability, development flexibility, and cost.

The system also enables direct recording of concurrent high-frequency stimulation waveforms, including kilohertz carrier signals used in temporal interference stimulation. Unlike many conventional EEG--tES platforms limited by lower sampling rates, the proposed system can capture both EEG activity and stimulation-related broadband components within a unified data stream. This capability may facilitate future artifact modeling, stimulation verification, and adaptive closed-loop neuromodulation studies \cite{care2024personalized}.


Several limitations should also be acknowledged. First, the present validation was performed using electrical bench tests and a gelatine head phantom rather than human participants. Although the phantom model provides a repeatable intermediate platform, future studies should evaluate system performance during real scalp recordings where motion artifacts, electrode impedance variability, and physiological noise are more complex. Second, the current platform supports eight EEG channels and two stimulation channels, which may be insufficient for some high-density neuroimaging or multi-focal stimulation applications. Third, although the system supports the hardware foundation for closed-loop neuromodulation, adaptive control algorithms were not implemented in the current study \cite{zhou2018toward}.

Future work will therefore focus on several directions. Expanding the number of sensing and stimulation channels would enable richer spatial monitoring and multi-site neuromodulation protocols. Investigations of tES artifact suppression and closed-loop neuromodulation strategies will further facilitate more precision interventions. Additional miniaturization, battery optimization, and wireless protocol enhancement may further improve suitability for long-term wearable use. Finally, human studies are needed to validate safety, comfort, and translational utility in practical and clinical scenarios.

Overall, the presented results suggest that modern low-power microcontrollers can provide sufficient computational capability for advanced bidirectional brain-interface systems when paired with efficient mixed-signal design. This may lower the barrier for future portable neuromodulation devices by complementing expensive and bulky laboratory architectures with more accessible embedded alternatives.

\section{Conclusions}

This paper presented a portable bidirectional EEG--tES platform that integrates multichannel neural recording and programmable transcranial electrical stimulation within a single MCU-based embedded system. The proposed architecture combines a high-resolution multichannel EEG front-end, high-speed data acquisition of up to 8~kHz per channel, and a custom DAC/DDS-based stimulation subsystem capable of generating multiple modalities including tDCS, tACS, and temporal interference stimulation. Experimental validation demonstrated high sensing fidelity, accurate stimulation output, and robust simultaneous EEG--tES acquisition without saturation or loss of critical waveform features. In particular, the system was able to directly capture concurrent high-frequency stimulation components, including kilohertz-range temporal interference carrier signals, which are not supported by many existing integrated EEG--tES platforms.

By delivering high performance within a compact, low-cost, and portable form factor, the proposed device provides a practical hardware foundation for next-generation closed-loop neuromodulation research. More broadly, the presented results demonstrate that efficient MCU-centered architectures, combined with companion host software, can serve as practical alternatives to bulky workstation-class or FPGA-centered systems for future wearable and real-world brain stimulation applications.


\FloatBarrier
\renewcommand*{\bibfont}{\small}
\printbibliography

\end{document}